&pdflatex

\documentclass[11pt,prd,aps,amssymb,amsmath,tightenlines,showpacs]{revtex4}
\usepackage{graphicx}

\usepackage[T1]{fontenc}
\usepackage[latin1]{inputenc}
\usepackage[english]{babel}
\usepackage{amsmath}
\usepackage{amsfonts}
\usepackage{amssymb}
\usepackage{color}

\def\cPT{\mathcal{PT}}

\begin{document}
\title{Analytic structure of eigenvalues of coupled quantum systems}

\author{Carl M. Bender$^1$}\email{cmb@wustl.edu}
\author{Alexander Felski$^2$}\email{alexander.felski@t-online.de}
\author{Nima Hassanpour$^1$}\email{nimahassanpourghady@wustl.edu}
\author{S. P. Klevansky$^{2,3}$}\email{spk@physik.uni-heidelberg.de}
\author{Alireza Beygi$^2$}\email{A.Beygi@ThPhys.Uni-Heidelberg.DE}

\affiliation{$^1$Department of Physics, Washington University, St. Louis,
Missouri 63130, USA\\
$^2$Institut f\"{u}r Theoretische Physik, Universit\"{a}t
Heidelberg, Philosophenweg 12, 69120 Heidelberg, Germany\\
$^3$Department of Physics, University of the Witwatersrand, Johannesburg,
South Africa} 

\begin{abstract}
By analytically continuing the coupling constant $g$ of a coupled quantum
theory, one can, at least in principle, arrive at a state whose energy is lower
than the ground state of the theory. The idea is to begin with the uncoupled
$g=0$ theory in its ground state, to analytically continue around an exceptional
point (square-root singularity) in the complex-coupling-constant plane, and
finally to return to the point $g=0$. In the course of this analytic
continuation, the uncoupled theory ends up in an unconventional state
whose energy is lower than the original ground state energy. However, it is
unclear whether one can use this analytic continuation to extract energy from
the conventional vacuum state; this process appears to be exothermic but one
must do work to vary the coupling constant $g$.
\end{abstract}
\maketitle

\section{Introduction}
\label{s1}
The analytic structure of eigenvalues of self-coupled systems, such as the
quantum anharmonic oscillator, has been studied in great depth. Singularities in
the coupling-constant plane have been identified as the cause of the divergence
of perturbation theory \cite{r1,r2}. These singularities are typically
square-root branch points and are associated with the phenomenon of level
crossing. These singularities are sometimes referred to as exceptional points
\cite{r3}. Studies of coupling-constant analyticity have revealed a remarkable
and generic phenomenon, namely, that the eigenvalues belonging to the spectrum
of the Hamiltonian are analytic continuations of one another as functions of the
complex coupling constant. Thus, the energy levels of a quantum system, which
are discrete when the coupling constant is real and positive, are actually
smooth continuations of one another in the complex-coupling-constant plane
\cite{r4}, and a simple geometric picture of quantization emerges: The discrete
eigenvalues of a quantum system are in one-to-one correspondence with the sheets
of the Riemann surface. The different energy levels of the Hamiltonian are
merely different branches of a multivalued energy function.

While this picture of quantization has emerged from studies of coupling-constant
singularities of self-coupled systems, this paper argues that an even more
elaborate picture arises from studies of coupled quantum systems. Consider, for
example, the simple case of two coupled quantum harmonic oscillators, one having
natural frequency $\nu>0$ and the other having natural frequency $\omega>0$. {\color{black} For definiteness, we assume that $\nu>\omega$.} The
Hamiltonian for such a system has the form
\begin{equation}
H=p^2+\nu^2x^2+q^2+\omega^2y^2+gxy,
\label{E1}
\end{equation}
where $g$ is the coupling parameter. For sufficiently large $|g|$ the
eigenvalues of $H$ become singular. To demonstrate this we rewrite the potential
$V(x,y)=\nu^2x^2+\omega^2y^2+gxy$ as
\begin{equation}
V(x,y)=\nu^2\left(x+\frac{gy}{2\nu^2}\right)^2+y^2\left(\omega^2-\frac{g^2}
{4\nu^2}\right).
\label{E2}
\end{equation}
We see immediately that on the line $x+gy/2\nu^2=0$ in the $(x,y)$ plane $V(x,
y)$ becomes unbounded below if $g^2>4\nu^2\omega^2$. Thus, while the potential
has a positive discrete spectrum when the coupling constant $g$ lies in the
range
\begin{equation}
-2\nu\omega<g<2\nu\omega,
\label{E3}
\end{equation}
we expect there to be singular points at $g=\pm2\nu\omega$ in the
coupling-constant plane. This result raises the question, What is the nature of
the singular points at $\pm2\nu\omega$?

Coupled-oscillator models have been studied in great detail in many papers
\cite{r5,r6,r7,r8,r9,r10,r11} and in particular for oscillator models of the
type in (\ref{E1}). The presence of singularities at $g=\pm2\nu\omega$ was noted
in \cite{r6}; however, the nature of singularities and the Riemann sheet
structure was not identified in any of these papers.

In this paper we show that the Riemann surface for the coupled-oscillator
Hamiltonian (\ref{E1}) consists of four sheets. The singularities at $g=\pm2\nu
\omega$ are square-root singularities, like the exceptional-point singularities
of self-coupled oscillators. However, if we cross either of the square-root
branch cuts, we enter a second sheet of the Riemann surface on which two {\it
new} square-root branch points appear. These new branch points are located at
$g=\pm i(\omega^2-\nu^2)$. If we cross either of the branch cuts emanating from
these new branch points, we enter a third sheet of the Riemann surface where
there are yet another pair of square-root singularities at $g=\pm2\nu\omega$,
unconnected with the singularities on sheets one and two. Crossing either of the
branch cuts emanating from these singularities at $g=\pm2\nu\omega$ takes us to
a fourth sheet of the Riemann surface. Not all energy levels of the coupled
harmonic oscillator mix among themselves as $g$ varies on this four-sheeted
Riemann surface. Rather, each energy level belongs to a quartet of energies that
are analytic continuations of one another. We find that the four sheets of the
Riemann surface correspond to four distinct {\it spectral phases} of the coupled
oscillator system (\ref{E1}).

We give a detailed description of these spectral phases in Sec.~\ref{s2}. We
explain below how such spectral phases arise. Let us consider a single harmonic
oscillator, whose dynamics are defined by the Hamiltonian
\begin{equation}
H=p^2+\nu^2x^2.
\label{E4}
\end{equation}
This simple quantum system actually has two spectral phases characterized by two
distinct spectra. To understand why, we assume that $\nu$ is a positive
parameter and we note that the $n$th eigenvalue $E_n$, which is defined by the
eigenvalue problem 
\begin{equation}
-\frac{d^2}{dx^2}\psi(x)+\nu^2x^2\psi(x)=E_n\psi(x)\quad(\psi\to0~{\rm as}~x\to
\pm\infty),
\label{E5}
\end{equation}
is given by
$$E_n=(2n+1)\nu\quad(n=0,1,2,3,...).$$
In \cite{r12} it was observed that if we analytically continue $\nu$ in a
semicircle in the complex-$\nu$ plane, that is, if we let $\nu=re^{i\phi}$ ($r$
real) and allow $\phi$ to run from $0$ to $\pi$, the eigenvalues change sign
even though the Hamiltonian remains unchanged. By this analytic continuation we
reach a new phase of the harmonic oscillator whose spectrum is {\it negative}
and {\it unbounded below}. Thus, the Hamiltonian (\ref{E4}) of the harmonic
oscillator has two distinct and independent real spectra that are related by
analytic continuation in the natural frequency $\nu$ of the oscillator.

How can one Hamiltonian (\ref{E4}) have two different spectra? The answer to
this question is that the positive spectrum is obtained by imposing the boundary
conditions in (\ref{E5}) in a pair of Stokes wedges \cite{r13,r14,r15,r4}
centered about the positive-real-$x$ and negative-real-$x$ axes. We refer to the
positive spectrum as the {\it conventional} one. These wedges have angular
opening $\pi/2$. The negative spectrum is defined by imposing the boundary
conditions in a pair of Stokes wedges containing and centered about the upper
and lower imaginary-$x$ axes. We refer to the negative spectrum as the {\it
unconventional} spectrum of the harmonic oscillator. These Stokes wedges also
have angular opening of $\pi/2$. To understand the configuration of the wedges
we examine the WKB geometrical-optics approximation \cite{r4}
\begin{equation}
\psi\sim e^{\pm\nu x^2/2}
\label{E6}
\end{equation}
to the solutions of the harmonic-oscillator eigenvalue equation (\ref{E4}). On
the basis of (\ref{E6}) we can see that the $90^\circ$ wedges in which the
eigenfunctions vanish rotate clockwise through an angle of $\pi/2$ as $\nu$
rotates anticlockwise through an angle of $\pi$. Thus, these two phases are
analytic continuations of one another and are analytically connected by
rotations in the complex-frequency plane.

A principal result of this paper is that, if we analytically continue the
physical system consisting of two coupled harmonic oscillators described by the
Hamiltonian in (\ref{E1}) in the coupling constant parameter $g$, we obtain all
four possibilities for the phases of the two oscillators in which each
oscillator is in a conventional or an unconventional phase. Thus, all four
phases are analytically connected on the Riemann surface of the complex coupling
constant $g$, even though the frequencies $\nu$ and $\omega$ are held fixed and
positive.

In Sec.~\ref{s2} we construct analytically the four-sheeted Riemann surface for
the coupled harmonic oscillator model (\ref{E1}). In Sec.~\ref{s3} we examine
the Riemann surface defined by the partition function for some zero-dimensional
quantum field theories. In general, the number of sheets in the complex Riemann
surface for these theories is smaller than the number of sheets for the
corresponding quantum-mechanical problem. For example, for the zero-dimensional
quantum field theory that is analogous to the quantum-mechanical oscillator
model of (\ref{E1}), the Riemann surface only has two sheets and not four
sheets. For a coupled pair of sextic models, the Riemann surface has six sheets,
indicating that this theory has six different phases. Section~\ref{s4} gives
some brief concluding remarks.

\section{Energy Levels of the Coupled Harmonic Oscillator}
\label{s2}
In this section we examine the analytic structure of the eigenvalues of the
coupled harmonic oscillator Hamiltonian (\ref{E1}). We begin by examining the
ground state, whose eigenfunction has the general form
\begin{equation}
\psi(x,y)=e^{-ax^2/2-by^2/2+cxy},
\label{E7}
\end{equation}
where $a$, $b$, and $c$ are constants to be determined. We substitute (\ref{E7})
into the eigenvalue equation $H\psi=E\psi$, which has the explicit form
\begin{equation}
-\psi_{xx}+\nu^2x^2\psi-\psi_{yy}+\omega^2y^2\psi+gxy\psi=E\psi.
\label{E8}
\end{equation}
We then equate the coefficients of $x^2$, $y^2$, $xy$, and $x^0y^0$ and obtain
the four equations
\begin{eqnarray}
\nu^2&=&a^2+c^2,\nonumber\\
\omega^2&=&b^2+c^2,\nonumber\\
0&=&2ac+2bc+g, \label{eq:9}\\
E&=&a+b. \label{eq:10}
\end{eqnarray}
Subtracting the first equation from the second and combining the result with the
third and fourth equations allows us to calculate $a$, $b$, and $c$, which we
then eliminate in favor of a single quartic polynomial equation for the
eigenvalue E:
\begin{equation}
E^4-2\left(\nu^2+\omega^2\right)E^2+\left(\nu^2-\omega^2\right)^2+g^2=0.
\label{E9}
\end{equation}
The solution to this equation involves nested square roots,
\begin{equation}
E(g)=\left[\nu^2+\omega^2+\left(4\nu^2\omega^2-g^2\right)^{1/2}\right]^{1/2},
\label{E10}
\end{equation}
and from this equation we see that $E(g)$ is a four-valued function of the
coupling constant $g$.

\begin{figure}[h!]
\begin{center}
\includegraphics[trim=0mm 5mm 0mm 5mm,clip=true,scale=0.30]{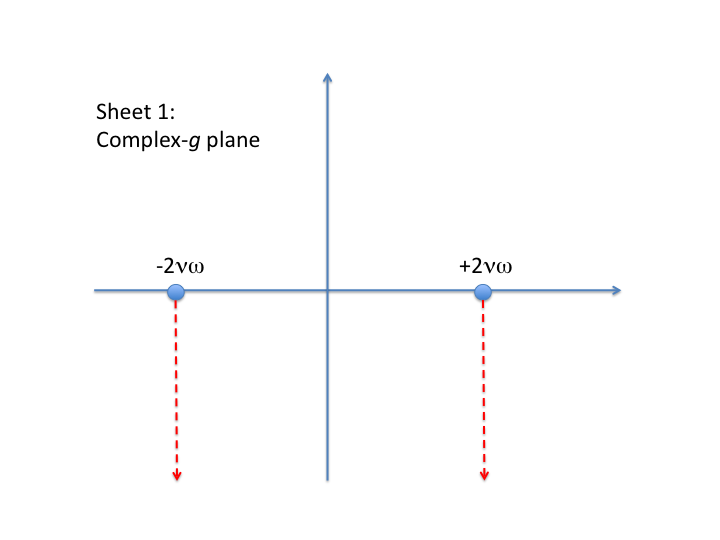}
\end{center}
\caption{[Color online] Sheet 1 of the complex Riemann surface of $E(g)$ in
(\ref{E10}). On this sheet both the inner and outer square roots are positive
when their arguments are positive. Branch points are indicated by blue dots
and branch cuts by red dashed lines. On this sheet $E(0)=\nu+\omega$.}
\label{F1}
\end{figure}

Let us make a grand tour of the Riemann surface on which $E(g)$ is defined. We
begin on Sheet 1, where both square-root functions are real and positive when
their arguments are real and positive. There are two obvious square-root branch
points (zeros of the inner square root) and these are located at
$g=\pm2\nu\omega$.
Square-root branch cuts emerge from each of these branch points and, as shown on
Fig.~\ref{F1}, we have chosen to draw these branch cuts as vertical lines going
downward. On Sheet 1
\begin{equation}
E(0)=\nu+\omega,
\label{E11}
\end{equation}
and because we assume that $\nu$ and $\omega$ are real and positive we see that
both oscillators are in their conventional ground states.

There are no other singularities on Sheet 1 that allow us to change the sign of
the outer square root. This is because at such a singular point the argument of
the outer square root function would have to vanish:
\begin{equation}
\nu^2+\omega^2+\sqrt{4\nu^2\omega^2-g^2}=0.
\label{E12}
\end{equation}
The solution of this equation is obtained by squaring $\nu^2+\omega^2=-\sqrt{4
\nu^2\omega^2-g^2}$:
\begin{equation}
-g^2=\left(\nu^2-\omega^2\right)^2,
\label{E13}
\end{equation}
so $-g^2$ is positive. The solution in (\ref{E13}) is {\it spurious} because
both terms in (\ref{E12}) are positive.

\begin{figure}[h!]
\begin{center}
\includegraphics[trim=0mm 5mm 0mm 5mm,clip=true,scale=0.30]{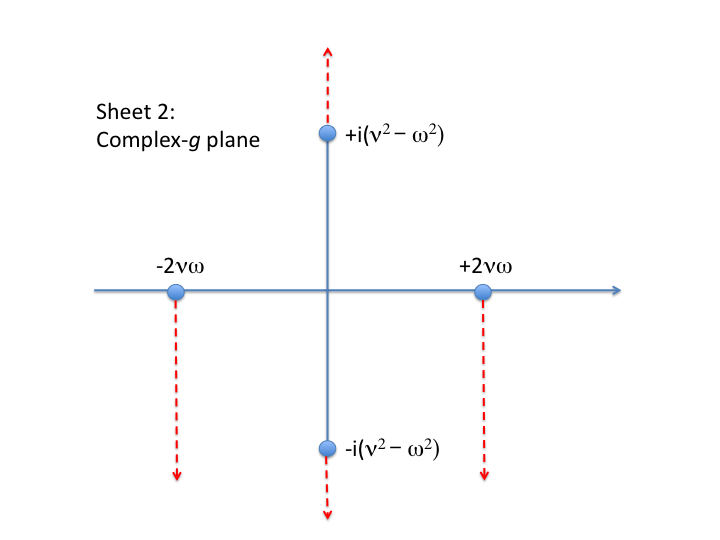}
\end{center}
\caption{Sheet 2 of the complex Riemann surface of $E(g)$ in (\ref{E10}). On
this sheet the inner square root in (\ref{E10}) is negative and the outer square
root is positive when their arguments are positive. On this sheet $E(0)$ is
$\nu-\omega$ (assuming that $\nu-\omega$ is positive).}
\label{F2}
\end{figure}

If we analytically continue $E(g)$ through either of the branch cuts on Sheet 1,
we arrive on Sheet 2, where the inner square root changes sign. Therefore, on
this sheet
\begin{equation}
E(0)=\nu-\omega,
\label{E14}
\end{equation}
assuming that $\nu>\omega$. Thus, the $x$ oscillator is in its conventional
ground state but the $y$ oscillator is in its unconventional ground state.
Because the inner square root returns negative
values when its argument is positive, the solution for $-g^2$ in (\ref{E13})
is {\it not} spurious. Therefore, there are new branch cuts associated with the
sign change of the outer square root; these branch cuts emanate from branch
points located at
\begin{equation}
g=\pm i\left(\nu^2-\omega^2\right).
\label{E15}
\end{equation}
All four branch cuts on Sheet 2 are shown on Fig. \ref{F2}. If we now pass
through a branch cut emanating from $\pm2\nu\omega$, we return to Sheet 1
but if we pass through a branch cut emanating from either branch point in 
(\ref{E15}), we enter Sheet 3.

\begin{figure}[h!]
\begin{center}
\includegraphics[trim=0mm 5mm 0mm 5mm,clip=true,scale=0.30]{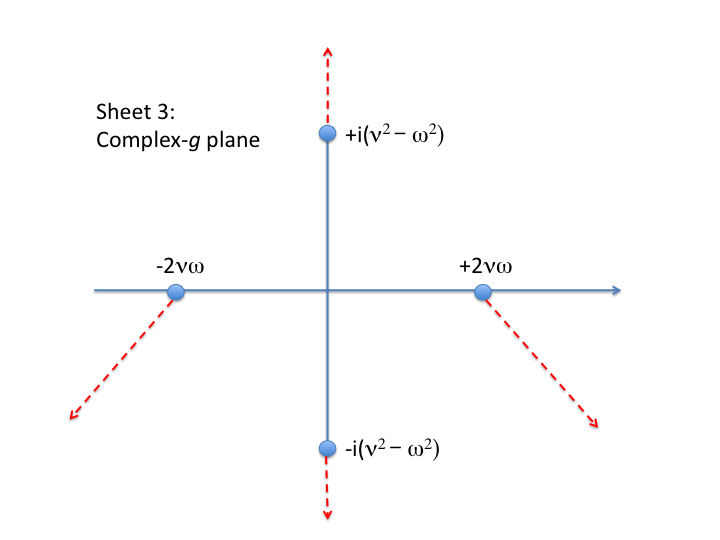}
\end{center}
\caption{Sheet 3 of the complex Riemann surface of $E(g)$ in (\ref{E10}). On
this sheet both the inner and outer square roots are negative when their
arguments are positive and thus $E(0)=-\nu+\omega$.}
\label{F3}
\end{figure}

On Sheet 3 there are two pairs of square-root branch cuts. The branch points
on the imaginary axis coincide with those on Sheet 2. However, there is a
new pair of branch points on the real axis at $g=\pm2\nu\omega$. Although
these branch points appear at the same locations as on Sheets 1 and 2, they
are unrelated to those branch points. We show this explicitly in Fig.
\ref{F3} by drawing the associated branch cuts differently. On this sheet both
the inner and outer square-root functions in (\ref{E10}) are negative and
\begin{equation}
E(0)=-\nu+\omega
\label{E16}
\end{equation}
when $\nu-\omega$ is positive. Now the $x$ oscillator is in an
unconventional ground state and the $y$ oscillator is in a conventional
ground state.

\begin{figure}[h!]
\begin{center}
\includegraphics[trim=0mm 5mm 0mm 5mm,clip=true,scale=0.30]{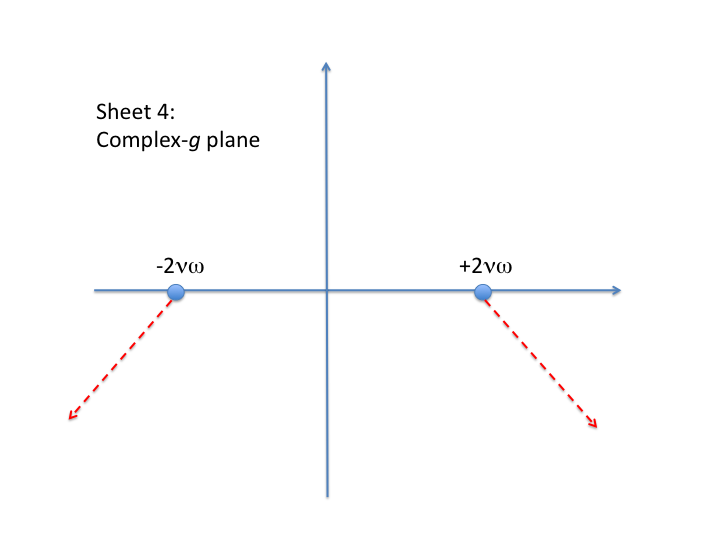}
\end{center}
\caption{Sheet 4 of the complex Riemann surface of $E(g)$ in (\ref{E10}). On
this sheet the inner square root is positive while the outer square root is
negative when their arguments are positive. On this sheet $E(0)=-\nu-\omega$.}
\label{F4}
\end{figure}

If we now pass through a branch cut emanating from (\ref{E15}), we return
from Sheet 3 to Sheet 2. However, if pass through a branch cut emanating from
$\pm2\nu\omega$, we enter Sheet 4. On this sheet there are only two branch
points, which are located at $\pm2\nu\omega$ (see Fig. \ref{F4}). On Sheet 4
\begin{equation}
E(0)=-\nu-\omega.
\label{E17}
\end{equation}
Both oscillators are now in unconventional ground
states.

To summarize, Figs. \ref{F1}-\ref{F4} describe each of the four branches of
the function $E(g)$ in (\ref{E10}). On these four branches $E(0)$ takes the
values given in (\ref{E11}), (\ref{E14}), (\ref{E16}), and (\ref{E17}). From
these four values of $E(0)$ we infer that by analytically continuing the
two-coupled-oscillator system in (\ref{E1}) through the entire Riemann
surface we access both phases, conventional and unconventional, of both
oscillators, even though the two frequency parameters $\nu$ and $\omega$
are held fixed.

\begin{figure}[h!]
\begin{center}
\includegraphics[trim=0mm 5mm 0mm 5mm,clip=true,scale=0.40]{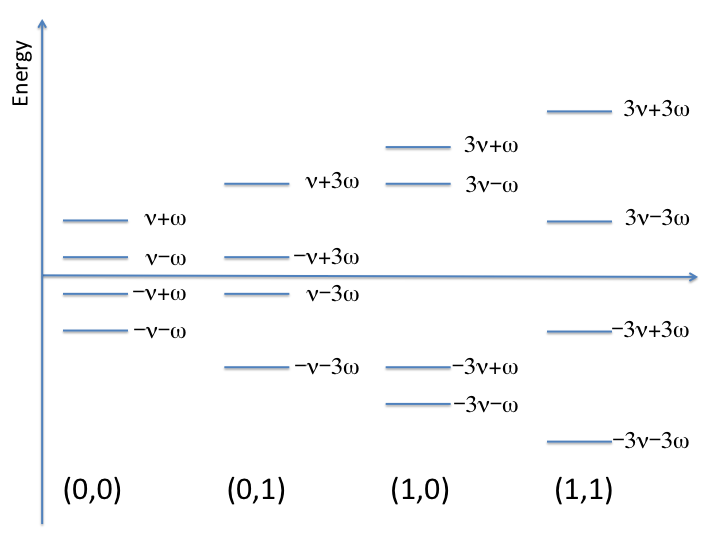}
\end{center}
\caption{First four quartets of energy levels associated with the Hamiltonian
(\ref{E1}). The quartets are labeled $(m,n)$, and the quartets shown are for 
$m=0,1$ and $n=0,1$. We have chosen the values $\nu=2$ and $\omega=1$ and
have plotted the values of $E(0)$ to scale. {\color{black} Note that each energy eigenvalue corresponds to the lowest such state on a different Riemann sheet.}}
\label{F5}
\end{figure}

The four-fold structure of the ground-state energy is repeated for all of the
energy levels. To verify this, we construct the eigenfunctions associated with
the other energy levels of the theory. These eigenfunctions consist of the
exponential in (\ref{E7}) multiplied by a polynomial $P(x,y)$. If $P(x,y)$ has the form
\begin{equation}
P(x,y) = Ax + By + Cxy +D,
\label{eq:18}
\end{equation}
the eigenvalue equation (\ref{E8}) leads to the three coupled equations (\ref{eq:9}) together with four alternatives for $E$:
\begin{eqnarray}
ED &=& (a+b)D, \label{eq:19a}\\
EA &=& A(3a+b) - 2Bc, \label{eq:19b}\\
EB &=& B(a+3b) -2Ac, \label{eq:19c}\\
EC &=& D(g+2bc+2ac) + 3C(a+b). \label {eq:19d}
\end{eqnarray}
For the quartet
of ground state energy levels described above, $D=1$, $A=B=C=0$, so that $P(x,y)=1$. We assign the label $(0,0)$ to this
quartet because it reduces to the (conventional and unconventional) ground
states of the $x$ and $y$ oscillators when $g=0$ and $c=0$. We use the designation $(0,1)$ for the quartet
$P(x,y)=y$,  $(1,0)$ for the quartet $P(x,y)=x$,  and $(1,1)$ for the quartet $P(x,y)
=xy$ that give rise to spectra in the decoupling limit $g=0$, $c=0$. In this limit, it follows again that $a^2=\nu^2$ and $b^2=\omega^2$, leading to four quartets with the additional three spectra arising from (\ref{eq:19b}) for (1,0) when $B=C=D=0$, (\ref{eq:19c}) for (0,1)  when $A=C=D=0$ and (\ref{eq:19d}) for (1,1) when $A=B=D=0$. These four quartets are illustrated in Fig. \ref{F5} for the
case $\nu=2$ and $\omega=1$. We emphasize that the energy levels of different
quartets are {\it not} analytic continuations of one another but the elements of
each quartet are analytic continuations of one another and branches of a
four-valued function defined on exactly the same the Riemann surface
pictured in Figs. \ref{F1}-\ref{F4}.

\section{Partition functions for zero-dimensional field theories}
\label{s3}
\subsection{Interacting quadratic field theory}
Let us examine the zero-dimensional field-theoretic equivalent of the
Hamiltonian (\ref{E1}). The partition function for this field theory
is given by the integral
\begin{equation}
Z(g) = \int\int dx\, dy\,e^{-\nu^2x^2-\omega^2y^2-gxy},
\label{E18}
\end{equation}
where both integration paths run from $-\infty$ to $\infty$.
We can evaluate the integral exactly by rearranging the
terms in the exponential as we did in (\ref{E2}):
\begin{equation}
Z(g)=\int\int dx\,dy\,e^{-\nu^2[x+gy/(2\nu^2)]^2-y^2[\omega^2-g^2/(4\nu^2)]}.
\label{E19}
\end{equation}
Simple transformations then reduce this to a product of two gaussian
integrals,
\begin{equation}
Z(g) = \int\int \frac{2\,dx\,dy}{\sqrt{4\nu^2\omega^2-g^2}}e^{-x^2-y^2},
\label{E20}
\end{equation}
which evaluate to 
\begin{equation}
Z(g)=\frac{2\pi}{\sqrt{4\nu^2\omega^2-g^2}}.
\label{E21}
\end{equation}

This partition function is a {\it double}-valued function of $g$ and is defined
on a two-sheeted Riemann surface. Like the coupled harmonic oscillator discussed
in Sec.~\ref{s2} {\color{black} the} square-root singularities {\color{black} are} located at $g=\pm2\nu\omega$.
However, unlike the case of the coupled harmonic oscillator, the Riemann surface
has two sheets and not four; these sheets correspond to the two possible
signs of $Z(g)$ and these two sheets correspond to the analogs of the
conventional-conventional theory and the unconventional-unconventional theory.
(To obtain the unconventional-unconventional theory from the
conventional-conventional theory we replace $x$ by $ix$ and $y$ by $iy$
and this changes the sign of the partition function.)
There is no analytic continuation to the partition function for a mixed
unconventional-conventional theory. This is because the path of integration is
included with the integral that defines the partition function. Given an
eigenvalue differential equation we are free to choose the boundary conditions
(we can require that the eigenfunctions vanish as $x\to\pm\infty$ or as $x\to\pm
i\infty$) but there is no such freedom in the case of an integral. To obtain
other phases we would have to change the path of integration in the definition
of the partition function.

We can generalize this calculation by including in the partition function
external fields $J$ and $K$ coupled to the $x$ and $y$ fields:
$$Z(J,K;g)=\int\int dx\,dy\,e^{-\nu^2x^2-\omega^2y^2-gxy+Jx+Ky}.$$
Evaluating this integral by following the same procedure as above, we now
find a more elaborate singularity structure,
$$Z(g)=\frac{2\pi}{\sqrt{4\nu^2\omega^2-g^2}}\exp\left(\frac{J^2\omega^2+K^2
\nu^2-gKJ}{4\omega^2\nu^2-g^2}\right),$$
which is again defined on a two-sheeted Riemann surface but in addition has
essential singularities at the square-root branch points. Consequently, all of
the Green's functions, which are obtained by taking derivatives with respect to
the external sources, have increasingly stronger singularities at $g=\pm2\nu
\omega$.

\subsection{Interacting sextic field theory}
A higher-power selfinteracting field theory that possesses a conventional real
spectrum and in addition possesses a real $\cPT$-symmetric spectrum has a sextic
interaction of the form $\phi^6$. We thus examine a field theory that describes
the coupling of two sextic oscillators and we choose a symmetric form for the
coupling. The partition function for the zero-dimensional version of this
coupled quantum field theory is
\begin{equation}
Z(g)=\int\int dx\,dy\,e^{-x^6-y^6-gx^3y^3}.
\label{E22}
\end{equation}
This sextic theory is more difficult to examine analytically. We begin by
expanding the coupling term as a series in powers of $g$:
\begin{equation}
Z(g)=\sum_{n=0}^\infty\frac{(-g)^n}{n!}\int\int dx\,dy\,e^{-x^6-y^6}x^{3n}y^{
3n}.
\label{E23}
\end{equation}

Since the $x$ and $y$ integrals run from $-\infty$ to $\infty$, only even
values of $n$ contribute to the partition function. When $n$ is even, we have
$$\int_{-\infty}^\infty dx x^{-x^6}x^{3n}=\frac{1}{3}\Gamma\left(\frac{n}{2}+
\frac{1}{6}\right),$$
but if $n$ is odd, the integral vanishes. Thus, we make the replacement $n=2m$
and re-express the partition function as a sum over $m$:
\begin{equation}
Z(g) = \frac{1}{9}\sum_{m=0}^\infty\frac{g^{2m}}{(2m)!}
\Gamma^2\left(m+\frac{1}{6}\right)
\label{E24}
\end{equation}
This sum is a hypergeometric series:
\begin{equation}
Z(g)=\frac{1}{9}\Gamma^2(1/6)\,\,_2F_1\left(\frac{1}{6},\frac{1}{6};\frac{1}{2};
\frac{g^2}{4}\right).
\label{E25}
\end{equation}

In general, the hypergeometric series has a radius of convergence of 1. (This is
easy to verify by using the Stirling approximation for the Gamma function.) This
implies that $Z(g)$ has a singularity on the circle $|g|=2$.

It is important to identify the precise location and nature of this singularity.
To do so we use the linear transformation formula \cite{r16}
\begin{eqnarray}
_2F_1(a,b;c;z)&=&\frac{\Gamma(c)\Gamma(c-a-b)}{\Gamma(c-a)\Gamma(c-b)}\,
_2F_1(a,b;a+b-c+1;1-z)\nonumber\\
&&\quad+(1-z)^{c-a-b}\frac{\Gamma(c)\Gamma(a+b-c)}{\Gamma(a)\Gamma(b)}\,
_2F_1(c-a,c-b;c-a-b+1;1-z).\nonumber
\end{eqnarray}
This transformation makes the singularity explicit because the hypergeometric 
function is analytic in the unit circle. Applying this transformation gives
\begin{eqnarray}
Z(g)&=&\frac{\sqrt{\pi}\,\Gamma^3(1/6)}{9\Gamma^2(1/3)}\,
_2F_1\left(\frac{1}{6},\frac{1}{6};\frac{5}{6};1-\frac{g^2}{4}\right)\nonumber\\
&&\quad+\left(1-\frac{g^2}{4}\right)^{1/6}\frac{\sqrt{\pi}\,\Gamma(-1/6)}{9}\,
_2F_1\left(\frac{1}{3},\frac{1}{3};\frac{7}{6};1-\frac{g^2}{4}\right),
\label{E26}
\end{eqnarray}
from which we conclude that $Z(g)$ is defined on a six-sheeted Riemann surface
and that the branch points on all six sheets of the Riemann surface are located
at $g=\pm2$, which corresponds with the singularities of the coupled harmonic
oscillator model at $\pm2\nu\omega$ with $\nu=\omega=1$.

More generally, we can examine the Green's functions $G_{\alpha,\beta}$ 
of the theory, which are defined as integrals of the form
\begin{equation}
G_{\alpha,\beta} \equiv \int\int dx\, dy\,x^\alpha y^\beta e^{-x^6-y^6-gx^3y^3},
\label{E27}
\end{equation}
where $\alpha$ and $\beta$ are integers. It is necessary that $\alpha+\beta$ is
even for the Green's function to be nonvanishing. Following the same analysis as
above, we find that all Green's functions are defined on a six-sheeted Riemann
surface and that the singularity in the complex-$g$ plane has the form
$$\left(1-\frac{g^2}{4}\right)^{(1-\alpha-\beta)/6}.$$
Thus, like the Green's functions for the coupled harmonic oscillator, we see
that the singularity becomes stronger with increasing $\alpha$ and $\beta$, but
the Green's functions are always six-valued functions of $g$.

\section{Conclusions}
\label{s4}
We have shown that a coupled quantum theory has a rich analytic structure as a
function of the coupling constant. By analytically continuing in the coupling
constant we can obtain different spectral phases of the {\it uncoupled} theory.
Indeed, if we think of the coupling constant as an external classical source,
then by varying this external source in a closed loop in the
complex-coupling-constant plane we can even imagine extracting energy from the
conventional ground state of such a theory, at least in principle. For example,
we can begin with the uncoupled harmonic-oscillator system (\ref{E1}) in its
conventional ground state (\ref{E11}). We then turn on the source $g$, smoothly
and continuously vary $g$, and finally turn off $g$ again when the system is in
the unconventional ground state (\ref{E17}). Such a process appears to be
exothermic because it extracts an amount of energy equal to $2\nu+2\omega$.
However, varying the coupling constant may require that we do work on the
system. {\color{black}Until now}, it is not clear what it means to vary a coupling constant
through complex values. {\color{black} However, remarkable progress on this is currently being made from an experimental point of view. It is experimentally possible to vary the
parameters of a system and by doing so to analytically continue from one energy
state to another. Such a process has actually been achieved in the laboratory by
smoothly varying the parameters of a microwave cavity \cite{r17} and, in doing
so, going continuously from one frequency mode to another. More recently, experiments have been performed in which an exceptional point is {\it dynamically} encircled \cite{r18,r19}. That is, a combination of physical parameters is varied in real time, and the system response is measured, allowing one to access different Riemann surfaces. While \cite{r19} emphasizes robust switching, \cite{r18} concerns itself with energy transfer between different states of a system, such as has been considered here in our illustrative prototypical system. An experimental approach, whether optomechanical or using micro, light, acoustic, or matter waves, may in the future yield experimental verification of the analytic continuation discussed in this work.}

{\color{black} Finally, these studies have been performed for {\it linear} couplings between the oscillators, which led to the four-fold structure shown here. It is to be expected that other types of couplings lead to different, possibly more complicated Riemann surfaces.}

\acknowledgments
CMB thanks the Graduate School of Heidelberg University for their hospitality.


\begin{thebibliography}{17}

\bibitem{r1} C.~M.~Bender and T.~T.~Wu, Phys.~Rev.~Lett.~{\bf 21}, 406 (1968).

\bibitem{r2} C.~M.~Bender and T.~T.~Wu, Phys.~Rev.~{\bf 184}, 1231 (1969).

\bibitem{r3} W.~D.~Heiss, J.~Phys.~A: Math.~Theor.~{\bf 45}, 444016 (2012).

\bibitem{r4} C.~M.~Bender and S.~A.~Orszag, {\it Advanced Mathematical
Methods for Scientists and Engineers} (McGraw-Hill, New York, 1976). 

\bibitem{r5} H.~Bateman, Phys.~Rev.~{\bf 38}, 815 (1931).

\bibitem{r6} C.~M.~Bender and H.~F.~Jones, J.~Phys.~A: Math.~Theor.~{\bf
41}, 244006 (2008).

\bibitem{r7} I.~L.~Aleiner, B.~L.~Altshuler, and Y.~G.~Rubo, Phys.~Rev.~B
{\bf 85}, 121301 (2012).

\bibitem{r8} C.~M.~Bender, M.~Gianfreda, B.~Peng, S.~K.~Ozdemir, and L.~Yang,
Phys. Rev. A {\bf 88}, 062111 (2013).

\bibitem{r9} C.~M.~Bender, M.~Gianfreda, and S.~P.~Klevansky, 
Phys.~Rev.~A {\bf 90}, 022114 (2014).

\bibitem{r10} I.~V.~Barashenkov and M.~Gianfreda, J.~Phys.~A:
Math.~Theor.~{\bf 47}, 282001 (2014).

\bibitem{r11} A.~Beygi, S.~P.~Klevansky, and C.~M.~Bender, Phys.~Rev.~A
{\bf 91}, 062101 (2015).

\bibitem{r12} C.~M.~Bender and A.~Turbiner, Phys.~Lett.~A {\bf 173}, 442
(1993).

\bibitem{r13} C.~M.~Bender and S.~Boettcher, Phys.~Rev.~Lett.~{\bf 80},
5243 (1998).

\bibitem{r14} S.~Schmidt and S.~P.~Klevansky, Phil.~Trans.~Roy.~Soc.~Lond.~{\bf
A371}, 20120049 (2013).

\bibitem{r15} C.~M.~Bender and S.~P.~Klevansky, Phys. Rev. Lett.~{\bf 105},
{\color{black} 031601} (2010).

\bibitem{r16} M.~Abramowitz and I.~A.~Stegun, {\it Handbook of Mathematical
Functions} (Dover, New York, 1972), Eq. 15.3.6.

\bibitem{r17} S.~Bittner, B.~Dietz, U.~G\"unther, H.~L.~Harney, M.~Miski-Oglu,
A.~Richter, and F.~Sch\"afer, Phys.~Rev.~Lett.~{\bf 108}, 024101 (2012).

{\color{black}
\bibitem{r18} H.~Xu, D. ~Mason, Luyao Jiang and J.~G.~E.~Harris, Nature {\bf 537}, 80 (2016).

\bibitem{r19} J.~Doppler, A.~A.~Mailybaev, J.~B\"ohm, U.~Kuhl, A.~Girschik, F.~Libisch, T.~J.~Milburn, P.~Rabl, N.~Moiseyev, and S.~Rotter, Nature, {\bf 537} 76 (2016).
}

\end{thebibliography}
\end{document}